\documentclass[aps,prb,twocolumn,superscriptaddress]{revtex4}

\usepackage{graphicx}
\begin{document}

%\preprint{}


\title{Neutron Diffuse Scattering from Polar Nanoregions in the
  Relaxor Pb(Mg$_{1/3}$Nb$_{2/3}$)O$_{3}$}
%
% PRB editors insist ``nano regions'' should be ``nanoregions.''
%



\author{K. Hirota}
\email[Corresponding author: ]{hirota@iiyo.phys.tohoku.ac.jp}
\affiliation{Department of Physics, Brookhaven National Laboratory, Upton,
New York 11973-5000}
\affiliation{Department of Physics, Tohoku University, Sendai 980-8578, Japan}

\author{Z.-G. Ye}
\affiliation{Department of Chemistry, Simon Fraser University,
Burnaby, British Columbia, Canada V5A~1S6}

\author{S. Wakimoto}
\affiliation{Department of Physics, University of Toronto, Toronto,
Ontario, Canada M5S~1A7}
\affiliation{Department of Physics, Brookhaven National Laboratory, Upton,
New York 11973-5000}

\author{P. M. Gehring}
\affiliation{NIST Center for Neutron Research, National Institute of
Standards and Technology, Gaithersburg, Maryland 20899-8562}

\author{G. Shirane}
\affiliation{Department of Physics, Brookhaven National Laboratory,
Upton, New York 11973-5000}

\date{\today}

%%
%% insert abstract here
%%

\begin{abstract}

%
% What we did.
%
We have studied the diffuse scattering in the relaxor
Pb(Mg$_{1/3}$Nb$_{2/3}$)O$_{3}$ (PMN) using triple-axis neutron scattering
techniques.  The diffuse scattering first appears around the Burns temperature
$T_{d} \approx 620$~K, indicating that its origin lies within the polar
nanoregions (PNR).  While the relative intensities of the diffuse scattering
around (101), (200), and (300) are consistent with those previously reported
by Vakhrushev {\it et al.}, they are, surprisingly, entirely different from
those of the lowest-energy transverse optic (TO) phonon.  This observation led
Naberezhnov {\it et al.} to claim that this TO mode could {\em not} be the
ferroelectric soft mode.  However, a recent neutron study by Gehring {\it et
al.} has unambiguously shown that the lowest-energy TO mode {\em does} soften
on cooling, and that the relative intensities are similar to those of
PbTiO$_{3}$.
%
% Draft 2
%
If the diffuse scattering in PMN originates from the condensation of a soft TO
mode, then the atomic displacements of the PNR {\em must} satisfy the center
of mass condition.  But, the atomic displacements determined from
%%
%% Rev.1 (begin)
%%
%% \textcolor{red}{(removed: x-ray)}
%%
%% Rev.1 (end)
%%
diffuse
scattering intensities do not fulfill this condition.  To resolve this
contradiction, we propose a simple model in which the total atomic
displacement consists of two components, $\delta_{CM}$ and $\delta_{shift}$. 
$\delta_{CM}$ is created by the soft mode condensation, and thus satisfies the
center of mass condition.  On the other hand, $\delta_{shift}$ represents a
uniform displacement of the PNR along their polar direction relative to the
surrounding (unpolarized) cubic matrix.
  Within the framework of this new model, we can successfully describe
  the neutron diffuse scattering intensities observed in PMN.

\end{abstract}



%
% \maketitle must follow title, authors, abstract, \pacs, and \keywords
%

\maketitle

%
% body of paper here
%

\section{Introduction}
\label{Introduction}

%
%  There is a slight contradiction here.  Relaxors show ``no phase 
%  transition into a ferroelectric state ...''  But PZN does become 
%  ferroelectric, and we classify it as a prototypical relaxor ...
%

Relaxors exhibit a broad maximum in the temperature
dependence of the dielectric constant and a significant frequency
dependence of the temperature $T_{max}$ at which this maximum occurs.
Although these peculiarities are often referred to as a ``diffuse''
phase transition, no macroscopic phase transition into a ferroelectric
state occurs at $T_{max}$.  Of particular interest are the lead-oxide
class of relaxors. 
%%
%% Rev.1 (begin)
%%
%% \textcolor{red}{
%%
This class displays extraordinary piezoelectric properties which persist over
a wide temperature range, thus presenting great appeal for device applications.
%%
%% }
%%
%% Rev.1 (end)
%%
Most of the lead-oxide relaxors are classified as B-site complex perovskite
compounds, in which an average B-site valence of $4+$ is realized by a
random occupancy of two different valence cations in a fixed ratio.
The prototypical lead-oxide relaxor system is
Pb$^{2+}$(Mg$^{2+}_{1/3}$Nb$^{5+}_{2/3}$)O$^{2-}_{3}$ (PMN) which,
along with Pb(Zn$_{1/3}$Nb$_{2/3}$)O$_{3}$ (PZN), shows an enormous
increase in piezoelectric character when doped with PbTiO$_{3}$.  In
spite of a decade of research, however, arguments over what intrinsic
mechanism drives the diffuse transition remain unsettled.

One of the most important concepts related to the microscopic
properties of relaxors is that of the so-called ``polar
nanoregions'' (PNR), the first experimental evidence for which was
obtained by Burns and Dacol.\cite{burns_1983} Through measurements of
the optic index of refraction $n(T)$ on single crystal specimens of
several disordered ferroelectric and relaxor compounds, including both
PMN and PZN, they observed that $n(T)$ deviates from a linear
temperature dependence at a temperature $T_{d}$ ($600-650$~K for PMN)
far above $T_{max}$ ($\approx 265$~K at 1~kHz for PMN).  They
proposed a model in which this unexpected high-temperature deviation
arises from small, randomly oriented, very local regions of
non-reversible polarization (the PNR) that begin to appear within the
otherwise non-polar crystal structure below $T_{d}$, which is often
called the Burns temperature.  Recent neutron inelastic scattering
studies on PZN and PZN doped with 8\% PbTiO$_{3}$ (PZN-8\%PT) in their
respective cubic phases at 500~K have shown that the lowest-energy
transverse optic (TO) phonon modes are overdamped for reduced wave
vectors $q$ less than a characteristic wave vector $q_{wf} \sim
0.2$~\AA$^{-1}$, but underdamped otherwise.\cite{gehring_2000} It is
now believed that this damping is caused by the PNR because they
couple strongly to the polar nature of the TO modes.

\begin{figure}[t]
\includegraphics{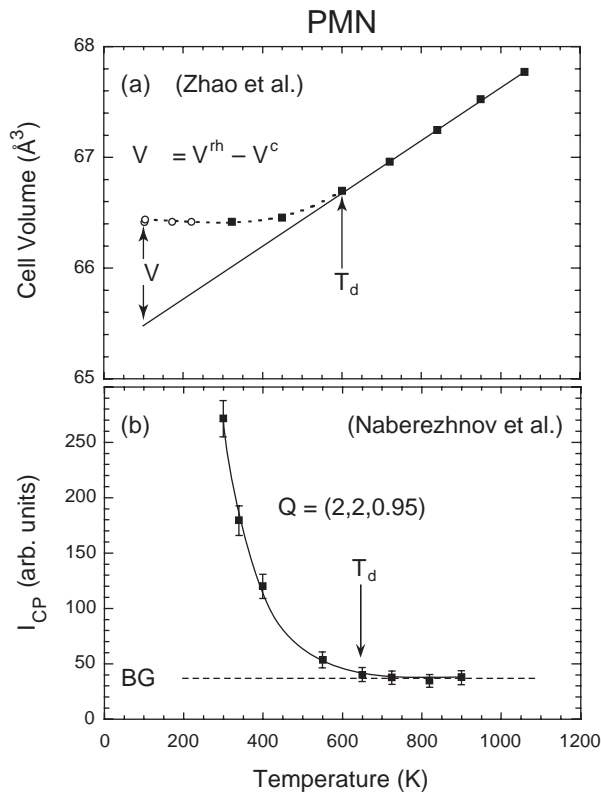}
\caption{(a) Temperature dependence of the unit cell volume, measured 
  by Zhao {\it et al.}\protect\cite{zhao_1998} (b) Temperature
  dependence of the neutron diffuse scattering intensity (central
  peak) at ${\bf Q}=(2, 2, 0.95)$, measured by Naberezhnov {\it et
    al.}\protect\cite{naberezhnov_1999} }
\label{FIG:Intro}
\end{figure}

Vakhrushev {\it et al.}\cite{vakhrushev_1989} carried out neutron
diffraction studies on PMN and observed strong diffuse scattering
which remains even at 500~K.  They found that the $q$-width of the
diffuse peak, which is inversely proportional to the correlation
length, is temperature dependent, and that both the Bragg and diffuse
peaks exhibit history dependent effects in the field-cooled (FC) and
zero field-cooled (ZFC) regimes, similar to that of a typical spin
glass.  The transition to this glass-like state occurs around 230~K,
which is slightly below $T_{max}$.  The diffuse scattering in PMN was
subsequently measured around 16 reciprocal lattice points, from which
the directions and relative magnitudes of the ionic displacements were
determined.\cite{vakhrushev_1995} The observed diffuse scattering is
broader along the direction transverse to the scattering vector ${\bf
  Q}$ (${\bf Q}_{\perp}$) than it is along the longitudinal direction
(${\bf Q}_{\parallel}$), a feature that is consistent with scattering
from ferroelectric fluctuations.  Bonneau {\it et
  al.}\cite{bonneau_1991} measured both x-ray and neutron powder
diffraction from PMN and reported that additional diffuse scattering
appears in the tails of the Bragg peaks below about 600~K.  Neutron
powder diffraction data taken by Zhao {\it et al.}\cite{zhao_1998}
show a marked break in the linear temperature dependence of the cubic
unit cell volume, also around 600~K, and is shown in
Fig.~\ref{FIG:Intro}(a).  More recently, Naberezhnov {\it et
  al.}\cite{naberezhnov_1999} have confirmed that the diffuse
scattering indeed becomes visible below 650~K, as shown in
Fig.~\ref{FIG:Intro}(b).  These results strongly imply that the
diffuse scattering in PMN results from the formation of the PNR at
$T_{d}$.

Naberezhnov {\it et al.}\cite{naberezhnov_1999} have also studied the
lattice dynamics of PMN using neutron inelastic scattering methods
between 300 and 900~K.  They calculated dynamical structure factors
using the room temperature diffuse scattering results of Vakhrushev
{\it et al.} \cite{vakhrushev_1989}, and concluded that (221) was the
best zone in which to look for the ferroelectric soft mode phonon.
However, they observed no underdamped soft TO phonon in the vicinity
of (221) even at 900~K.  Instead, they observed well-shaped peaks from
the transverse acoustic (TA) and the lowest-lying TO modes near (220),
although nearly no diffuse scattering was observed around (220).  They
therefore concluded that the observed lowest-lying TO mode could not
be identified with the ferroelectric soft mode because it exhibited a
structure factor that was absolutely inconsistent with that expected
from the ferroelectric diffuse scattering peak intensities.  However,
a recent neutron inelastic scattering study by Gehring {\it et
  al.}\cite{gehring_2001} has unambiguously shown that the
lowest-energy zone center TO mode does soften significantly on cooling
from 1100~K to $T_{d}$, below which it becomes overdamped, and that
the relative phonon intensities measured in different zones are very
similar to those found in PbTiO$_{3}$, a prototypical (displacive)
ferroelectric system.

It is thus clear that an important discrepancy exists between the
relative intensities of the soft TO mode and those of the diffuse
scattering in PMN.  Since each is related to either dynamic or static
atomic displacements toward the same ferroelectric state, the relative
intensities should be consistent.  The aim of this paper is to
determine how the diffuse scattering in PMN connects to the soft TO
mode, and thereby resolve this discrepancy.  To study the diffuse
scattering in PMN in detail, we have carried out neutron diffraction
measurements from 200 to 700~K.  We have concentrated on the diffuse
scattering around the three Bragg reflections (101), (200), and (300),
and measured the $q$-profiles and the temperature dependence.  These
results, as well as the experimental conditions, are summarized in
Section~\ref{Diffuse}.  In Section~\ref{Phonon} we describe our phonon
structure factor calculations using the atomic displacements
determined from the diffuse scattering intensities, which helps to
clarify the inconsistency between the relative intensities of the soft
TO mode and of the diffuse scattering.  We believe this discrepancy
can only be resolved if one assumes the simple model in which the PNR
are shifted along their polar direction relative to the surrounding
cubic matrix.  Finally, we compare our experimental results with those
from previous neutron and x-ray scattering studies, and discuss one
possible origin of this new concept, which we call a ``phase-shifted
condensed soft mode,'' in Section~\ref{Discussion}.

\section{Diffuse Scattering}
\label{Diffuse}

The neutron diffuse scattering data presented here were obtained on
the BT9 triple-axis spectrometer located at the NIST Center for
Neutron Research.  The data were taken with fixed incident neutron
energy $E_{i}=14.7$~meV ($\lambda=2.36$~\AA), and with horizontal beam
collimations 40$'$-46$'$-S-40$'$-80$'$ or 40$'$-10$'$-S-10$'$-80$'$
(``S'' = sample).  Single crystals of PMN were grown by a top-seeded
solution growth technique using PbO as flux.  The growth conditions
were determined based on the pseudo-binary phase diagram established
from PMN and PbO.\cite{ye_1990} An as-grown single crystal with a
nearly half-cubic morphology, having 
%% 
%% Rev.1 (begin)
%%
%% \textcolor{red}{
%%
%%
%% dimensions $\times$(0.8~cm$\times$0.9~cm$\times$0.9~cm) = 0.4 cm$^3$
a volume of 0.4 cm$^{3}$
%%
%% }
%%
%% Rev.1 (end)
%%
and a weight of 3.25~g, was used for the diffuse scattering measurements.
The crystal exhibits three naturally grown $\{1\ 0\ 0\}_{cubic}$ facets,
and was mounted on a boron nitride post using tantalum wire with one
of the facets facing vertically, and then attached to the cold-head of
a high-temperature closed-cycle helium refrigerator.  This orientation
gave access to reflections of the form $(h0l)$ in the scattering
plane.  The lattice constant of PMN is $a=4.04$~\AA\ at room
temperature, thus 1~rlu (reciprocal lattice unit) corresponds to
$2\pi/a=1.553$~\AA$^{-1}$.

\begin{figure}[t]
\includegraphics{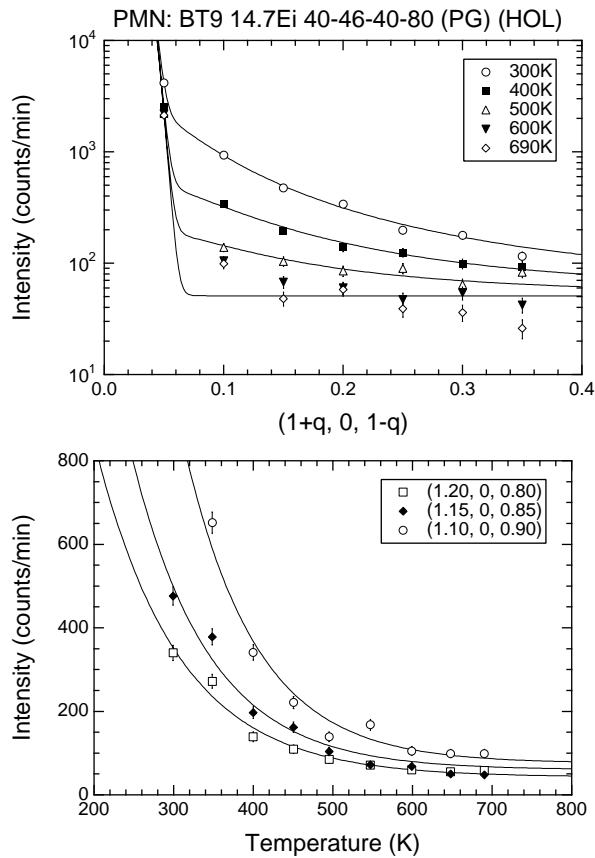}
\caption{(a) Diffuse scattering profiles around (101) along $[10\bar{1}]$ at
  300, 400, 500, 600, and 690~K.  Curves are fits to a combination of
  Gaussian (Bragg peak) and Lorentzian (diffuse scattering)
  lineshapes.  (b) Temperature dependence of the diffuse scattering
  intensities at (1.20, 0, 0.80), (1.15, 0, 0.85) and (1.10, 0, 0.90).
  The diffuse scattering clearly begins increasing around $T_{d}$.}
\label{FIG:Diffuse-Tdep}
\end{figure}

Figure~\ref{FIG:Diffuse-Tdep}(a) shows peak profiles of the transverse
diffuse scattering in the vicinity of (101).  Note that the $y$-axis
is displayed on a logarithmic scale.  At 690~K, there is no diffuse
scattering.  The peak profile is well-described by a Gaussian function
with a full-width at half-maximum (FWHM) corresponding to that of the
instrumental $q$-resolution at the (101) Bragg peak.  With decreasing
temperature, a weak signal starts to emerge out of the (101) Bragg
peak, growing more rapidly as the temperature is lowered.  The
observed diffuse scattering can be nicely fit with a Lorentzian
function as shown in Fig.~\ref{FIG:Diffuse-Tdep}(a).  The temperature
dependence of the diffuse scattering is more clearly presented in
Fig.~\ref{FIG:Diffuse-Tdep}(b).  The diffuse scattering appears below
about $600-650$~K, consistent with the Burns temperature $T_{d} \sim
620$~K for PMN, and increases almost exponentially with decreasing
temperature.

\begin{figure}[t]
\includegraphics{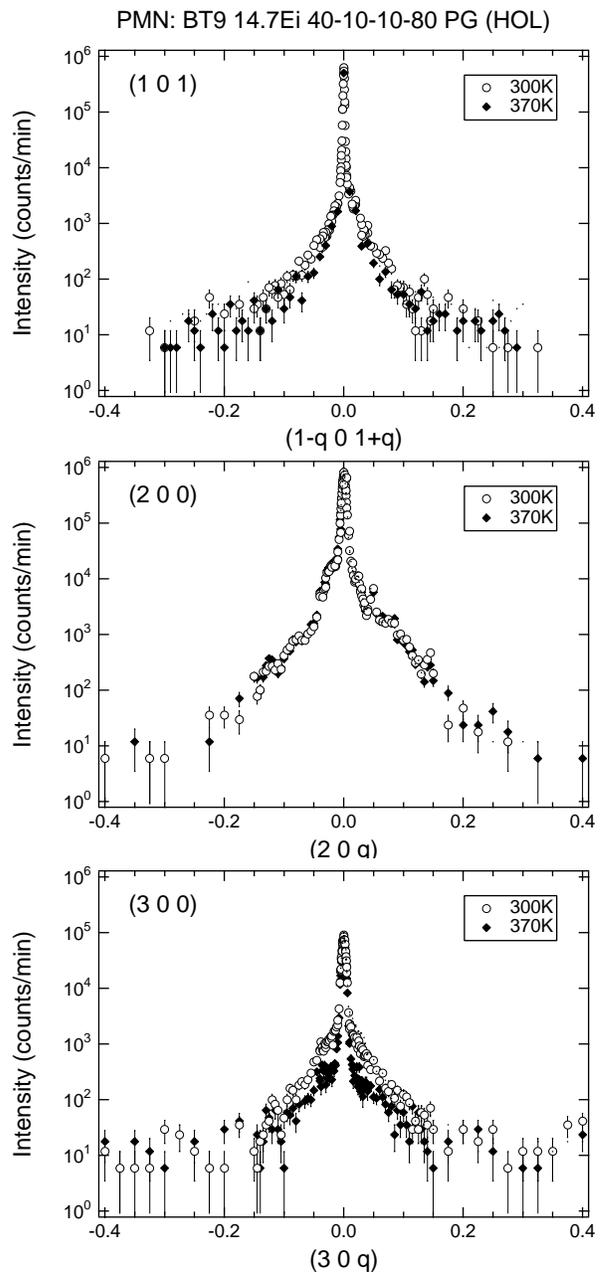}
\caption{Profiles of the diffuse scattering peaks at (a) $(1-q,0,1+q)$, 
  (b) $(2,0,q)$, and (c) $(3,0,q)$ at $T=300$ and 370~K.}
\label{FIG:Diffuse-Profiles}
\end{figure}

\begin{figure}[t]
\includegraphics{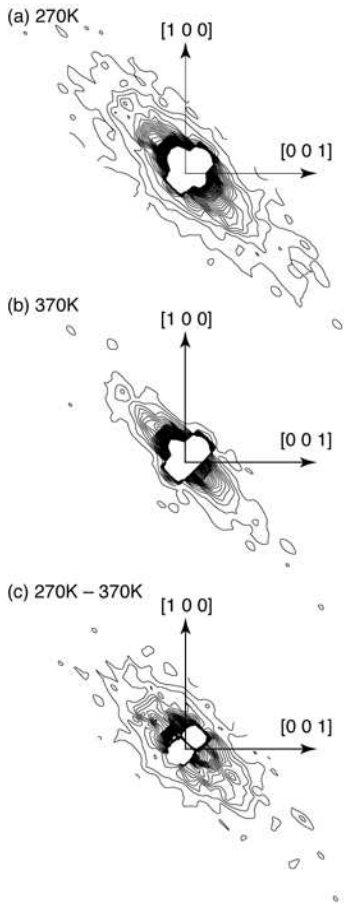}
\caption{Intensity contours of the diffuse scattering around (101) 
  at (a) 270~K and (b) 370~K.  (c) Difference between (a) and (b).}
\label{FIG:Diffuse-Contour}
\end{figure}

The diffuse scattering peak profiles were more closely examined by
employing a tighter collimation of 40$'$-10$'$-S-10$'$-80$'$.
Figure~\ref{FIG:Diffuse-Profiles} depicts the transverse diffuse
scattering profiles at (101), (200), and (300) at 300 and 370~K.
Although the difference in the diffuse scattering intensity between
these two temperatures is not very large, as is apparent from
Fig.~\ref{FIG:Diffuse-Tdep}, it is nevertheless discernable at (101).
A significantly larger difference is observed at (300), which
indicates that the diffuse scattering is stronger at (300) than at
(101).  No difference in intensity at finite $q$ near (200) was found,
which suggests the diffuse scattering is very weak at (200).  We
speculate that the tail-like feature near $q=0$ for (200) is due to
contamination from the TA phonon, which is quite strong near (200), as
some TA mode scattering will inevitably spill into the elastic profile
because of the imperfect instrumental energy resolution.  The relative
intensities of the (101), (200), and (300) diffuse scattering peaks
are consistent with those reported in the previous neutron diffraction
measurement by Vakhrushev {\it et al.}\cite{vakhrushev_1989} We have
also carried out a detailed survey of the (101) diffuse scattering,
the results of which are shown in Fig.~\ref{FIG:Diffuse-Contour}.  The
diffuse scattering is highly elongated along the direction transverse
to the scattering vector, which is also consistent with the previous
report.\cite{vakhrushev_1989} Similar neutron diffuse scattering
results at (101) and (200) were recently observed by Koo {\it et al.}
in a single crystal sample of PMN doped with 20\%
PbTiO$_3$.~\cite{koo_pmn20}

\section{Ferroelectric Phonon}
\label{Phonon}

Gehring {\it et al.}\cite{gehring_2001} have recently reported that
the lowest-energy zone center TO mode in PMN softens on cooling from
1100~K to $T_{d}$, and that the relative intensities measured in
different zones are very similar to those found in PbTiO$_{3}$.  This
result sharply contradicts the fact that the relative intensities of
the diffuse scattering in different zones are entirely different from
those of the lowest-energy transverse optic (TO) phonons, an
observation which led Naberezhnov {\it et al.}\cite{naberezhnov_1999}
to claim that the lowest-energy TO mode could {\em not} be the
ferroelectric soft mode.
%%
%% Rev.2 (begin)
%%
To understand how PMN undergoes the {\em diffuse transition} through the
formation of the PNR, this contradiction must be resolved.
%%
%% Rev.2 (end)
%%
In this section, we revisit the soft mode of PMN and introduce a new concept,
the ``phase-shifted condensed soft mode,'' as a microscopic description of PNR
formation.

\begin{figure}[t]
\includegraphics{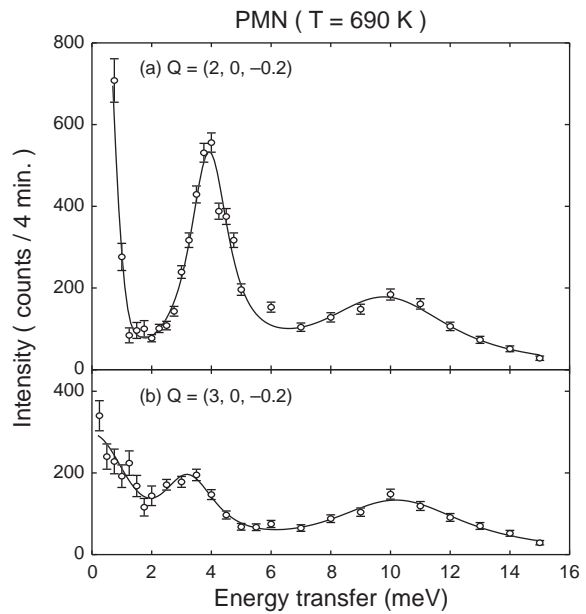}
\caption{Energy spectra of the transverse acoustic and optic phonons at
  690~K, i.e., above the Burns temperature $T_{d}$, around (a) (200)
  and (b) (300).}
\label{FIG:Phonon-Profiles}
\end{figure}

As shown in the previous section, we observe no diffuse scattering
around (200), consistent with previous reports.  However, TO phonons
are clearly observed around both (200) and (300) as shown in
Fig.~\ref{FIG:Phonon-Profiles}, which shows two constant-{\bf Q} scans
at 690~K (i.e., above the Burns temperature $T_{d}$) at $(2, 0, -0.2)$
and $(3, 0, -0.2)$.  The solid curves are fits to two Lorentzian
functions of $E$, the neutron energy transfer, convolved with the
proper instrumental resolution function.  From the fitting results,
the ratio of the two TO phonon intensities is
\begin{equation}
  \frac{|F_{obs}(200)|^{2}}{|F_{obs}(300)|^{2}}=1.24\pm0.20.
\label{EQ:Ratio-Exp}
\end{equation}
According to the pioneering work by Harada {\it et al.}\cite{harada_1970},
which describes the determination of the normal mode vibrational displacements
in perovskites from measured phonon intensities, the relative intensities of
phonons are determined mostly by the ratio of two dominant modes, i.e., the
Slater mode and the Last mode.  In the Slater mode, the oxygen and Mg/Nb (MN) atoms
vibrate in opposition while the Pb atoms remain stationary.  The Last mode
corresponds to opposing motions of the (Mg/Nb)O$_{6}$ octahedra and
the Pb atoms.  In both modes, the three oxygen atoms in the unit cell
move as a rigid unit.  We define the ratio of the two modes as
$S=S_{2}/S_{1}$, where $S_{1}$ and $S_{2}$ represent the contribution from the
Slater mode (${\bf s}_{1}$) and Last mode (${\bf s}_{2}$), respectively.

\begin{figure}[t]
\includegraphics{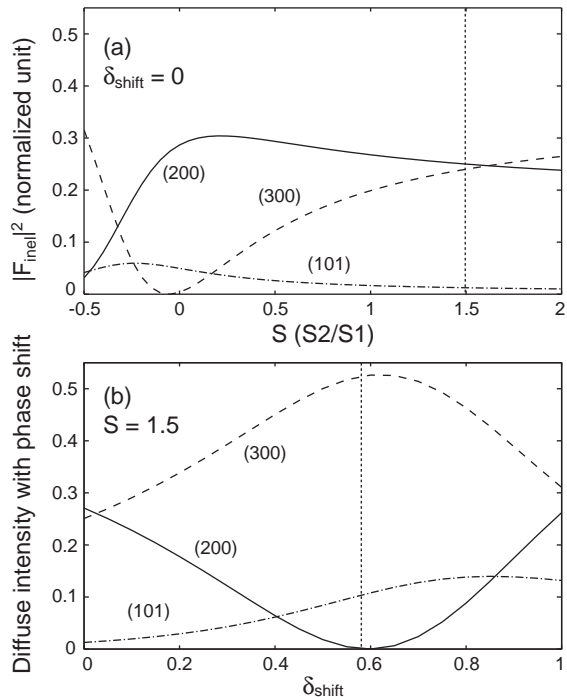}
\caption{(a) Calculated intensities of the TO phonon mode at (101), 
  (200) and (300) as a function of the ratio $S$ between the $S_{1}$
  (Slater) and $S_{2}$ (Last) modes.  
%%
%% Rev.1 (begin)
%%
%% \textcolor{red}{(Removed: Inset ...)}
%%
%% Rev.1 (end)
%%
  (b) Calculated
  intensities of the TO phonon mode at (101), (200) and (300) as a
  function of the phase shift of the PNR along their polar direction
  relative to the surrounding cubic matrix.}
\label{FIG:Phonon-Calc}
\end{figure}

Let us first determine the ratio $S$.  Figure~\ref{FIG:Phonon-Calc}(a)
shows the inelastic structure factors $|F_{inel}|^{2}$ at (101), (200)
and (300) as a function of $S$, which are calculated using the formula
given by Harada {\it et al.}\cite{harada_1970} In this calculation, we
assume that the atomic displacements
%%
%% Rev.1 (begin)
%%
%% \textcolor{red}{
%%
${\bf s}_{1}$ and ${\bf s}_{2}$
are parallel to the [100] direction for the sake of simplicity.
Combining Eqs.(1) and (5) of Ref.~\onlinecite{harada_1970}, we have obtained
$|F_{inel}|^{2}$ for the lowest branch at the zone center $({\bf q}=0)$ using
constant-{\bf Q} scans as follows:
\begin{equation}
\left|F_{inel}\right|^{2} =
\sum_{\kappa}\left[{\bf Q}\cdot{\bf\xi}_{\kappa}\right]
b_{\kappa}
\exp\left({-W_{\kappa}}\right)
\exp\left({i{\bf G}\cdot{\bf R}_{\kappa}}\right),
\label{EQ:F_inel}
\end{equation}
where the scattering vector (${\bf Q}$) is the sum of the reciprocal lattice
vector (${\bf G}$) and the phonon wave number (${\bf q}=0$), and the
normalized displacement vector for the $\kappa$-th atom, ${\bf\xi}_{\kappa}$,
in our condition is
\begin{equation}
\xi_{\kappa} = {\bf s}_{\kappa 1}+S\ {\bf s}_{\kappa 2}.
 \label{EQ:Xi}
\end{equation} 
From Table~1 of Ref.~\onlinecite{harada_1970}, the atomic displacements along
the [100] direction are 
\begin{eqnarray}
{\bf \xi}_{\rm Pb} & = & -Sk' \nonumber \\
{\bf \xi}_{\rm MN} & = & -k + S \nonumber \\
{\bf \xi}_{\rm O}  & = & 1 + S,
\label{EQ:Displacements}
\end{eqnarray}
where the center of mass condition requires $k=3M(\rm O)/M(\rm MN)=0.686$ and
$k'=[M({\rm MN})+3M({\rm O})]/M({\rm Pb}) = 0.596$. 
The calculated values are normalized by the sum of $|F_{inel}|^{2}$ over all
reflections that can be reached with an incident neutron energy $E_i =
14.7$~meV: (100), (110), (200), (210), (220), (300), (310) and their
equivalent.  Debye-Waller factors,
$\exp\left(-W_{\kappa}\right)=\exp(-B_{\kappa}|{\bf Q}/{4\pi}|^{2})$, are
calculated using $B_{Pb}=0.915$ and
$B_{NM}=0.483$ and $B_{O}=1.02$.\cite{verbaere_1992}
%%
%% }
%%
%% Rev.1 (end)
%%
Figure~\ref{FIG:Phonon-Calc}(a) indicates that
the experimentally observed intensity ratio of
$|F_{obs}(200)|^2/|F_{obs}(300)|^2$ = 1.24 obtained in
Eq.~(\ref{EQ:Ratio-Exp}) is realized
%%
%% Rev.1 (begin)
%%
%% \textcolor{red}{
%%
between $S=1.0$ and $1.5$.
%%
%% }
%%
%% Rev.1 (end)
%%

If the diffuse scattering of PMN originates from the condensation of
the soft TO mode, then
%%
%% Rev.2 (begin)
%%
the above formula should be also applicable to the diffuse scattering
intensities and that 
%%
%% Rev.2 (end)
%%
the atomic displacements determined from the diffuse scattering intensities
must also satisfy the center of mass condition:
\begin{equation}
  \sum_{\kappa}\delta(\kappa)M(\kappa)=0,
\label{EQ:Center-of-Mass}
\end{equation}
where $\delta(\kappa)$ and $M(\kappa)$ are the displacement and the atomic mass
of the $\kappa$-th atom, respectively.  However, the values $\delta(\kappa)$
determined from the diffuse scattering intensities by Vakhrushev {\it et
al.}\cite{vakhrushev_1995},
\begin{equation}
\begin{array}{rclrcl}
\delta({\rm Pb}) & = &  1.00, & M({\rm Pb}) & = & 207, \\
\delta{(\rm MN}) & = &  0.18, & M({\rm MN}) & = & 70, \\
\delta({\rm O})  & = & -0.64, & M({\rm O})  & = & 16, 
\label{EQ:delta-Vakhrushev}
\end{array}
\end{equation}
do {\em not} satisfy the center of mass condition.  This contradiction
is a natural corollary to the observed inconsistency between the
relative intensities of the diffuse and the soft TO mode scattering.

To resolve this discrepancy, we propose a simple model in which the
atomic displacements consist of two components,
$\delta(\kappa)=\delta_{CM}(\kappa)+\delta_{shift}$.  In this model,
$\delta_{CM}$ is induced by the condensation of the soft mode, and
thus satisfies the center of mass condition.  On the other hand,
$\delta_{shift}$ shifts the PNR along their polar directions relative
to the surrounding cubic matrix, which is, in other words, a phase
shift.  By applying this model, the $\delta(\kappa)$ values listed in
Eq.~(\ref{EQ:delta-Vakhrushev}) can be separated into two parts;
\begin{eqnarray}
\delta_{CM}({\rm Pb}) & = & 0.42,  \nonumber \\
\delta_{CM}{(\rm MN}) & = & -0.40, \nonumber \\
\delta_{CM}({\rm O})  & = & -1.22,
\label{EQ:delta-COM}
\end{eqnarray}
and
\begin{equation}
\delta_{shift}=0.58.
\label{EQ:phase-shift}
\end{equation}
Note that the $\delta_{CM}$ values are uniquely determined from the
center of mass condition and can be further decomposed into
contributions from the Slater mode and the Last mode.
%using the following conditions;
%%
%%
%\begin{eqnarray}
%{\rm Slater\ mode}:\
%\frac{\delta_{CM}({\rm MN})}{\delta_{CM}({\rm O})}
%& = & \frac{-3M({\rm O})}{M({\rm MN})}\\
%{\rm Last\ mode}:\
%\frac{\delta_{CM}({\rm MN})}{\delta_{CM}({\rm Pb})}
%& = & \frac{\delta_{CM}({\rm MN})}{\delta_{CM}({\rm O})} \nonumber \\
%& = &  \frac{M({\rm MN})+3M({\rm O})}{M({\rm Pb})}.
%\end{eqnarray}
The corresponding ratio $S$ between the Slater mode and Last mode
contributions is 1.5, which is consistent with the value obtained in
Fig.~\ref{FIG:Phonon-Calc}(a).  Note that the phase shift
$\delta_{shift}$ becomes effective only when the PNR condense out from
the soft TO mode.  Hence the phonon intensities are determined only by
the $\delta_{CM}$ values.

Now that we have confirmed that the $\delta_{CM}$ values derived from
the diffuse scattering intensities are consistent with the inelastic
scattering intensities of the soft TO mode, we can calculate how the
diffuse scattering intensities depend on the phase factor
$\delta_{shift}$ with the constraint $S=1.5$ using the $\delta_{CM}$
values listed in Eq.~(\ref{EQ:delta-COM}).  The results are shown in
Fig.~\ref{FIG:Phonon-Calc}(b).  Note that the values at
$\delta_{shift}=0$ are identical to those at $S=1.5$ in
Fig.~\ref{FIG:Phonon-Calc}(a) as expected.  At $\delta_{shift}=0.58$,
which was derived above to satisfy the center of mass condition, the
diffuse intensity around (200) approaches zero, in agreement with
experiment.  Thus, our ``phase-shifted condensed soft mode'' model
consistently explains both the diffuse scattering and the soft mode
intensities from the same atomic displacements.

\begin{figure}[t]
\includegraphics{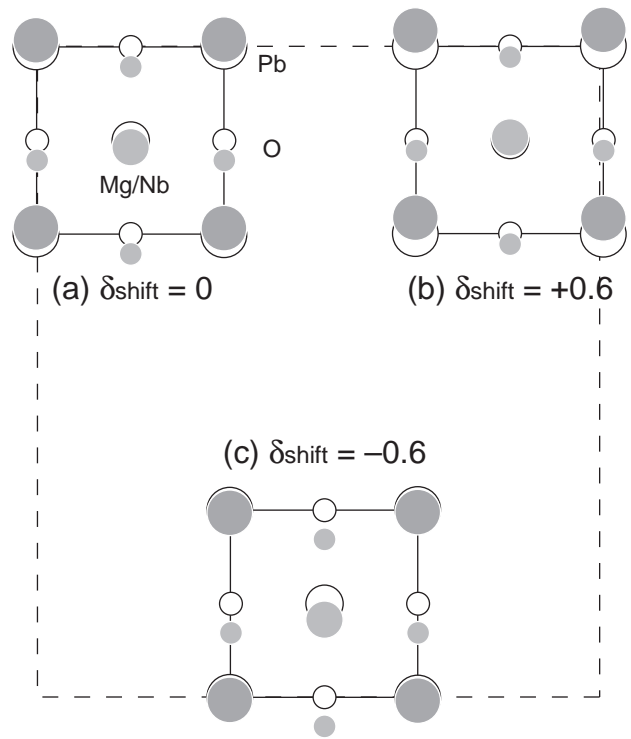}
\caption{Atomic displacements corresponding to three different phase 
  shift values; (a) $\delta_{shift}=0$, (b) $\delta_{shift}=0.6$ and
  (c) $\delta_{shift}=-0.6$.  The open circles represent the normal
  atomic positions without displacement, whereas the shaded circles
  show the atomic positions with both displacement and shift.}
\label{FIG:Lattice} 
\end{figure}

Lastly, we consider the phase shift from the crystal lattice point of
view.  Figure~\ref{FIG:Lattice} depicts the atomic displacements
schematically with three different phase shift values.
Figure~\ref{FIG:Lattice}(a) shows the displacements of
Eq.~(\ref{EQ:delta-COM}) without the phase shift, which satisfies the
center of mass condition and gives the same relative intensities for
both the diffuse scattering and the soft TO mode.  However, once all
the atoms undergo an additional shift along the same direction by the
same amount $\delta_{shift} = +0.6$, the center of mass condition is
no longer satisfied, resulting in the {\em accidental} disappearance
of the (200) diffuse neutron scattering intensity.
Figure~\ref{FIG:Lattice}(c), which depicts a uniform shift of
$\delta_{shift} = -0.6$, is shown just for comparison.

\section{Discussion}
\label{Discussion}

In the preceding sections we have demonstrated how the inconsistency
between the diffuse scattering and the soft TO mode intensities is
explained naturally within the framework of a ``phase-shifted
condensed soft mode.''  Our data on the diffuse scattering
cross-sections in three different Brillouin zones along with our
measurements of the soft TO mode phonons in two different zones serve
as a confirmation of the detailed measurements already published by
other groups.
\cite{vakhrushev_1989,vakhrushev_1995,bonneau_1991,naberezhnov_1999}
But while our data are in agreement, our interpretations differ
entirely.  The simplicity of our model argues in its favor.  Other
than the standard picture of condensed local ferroelectric
fluctuations created by a soft phonon, there is only one added
feature, that being a uniform phase shift of the PNR relative to
the surrounding cubic matrix.  We believe the simpler solution is the
correct one.

The process of separating the observed atomic displacements into optic
mode displacements $\delta_{CM}$ and a single phase shift parameter
$\delta_{shift}$ is straightforward and unique, and requires no
adjustable parameters.
%%
%% Rev.1 (begin)
%%
%% \textcolor{red}{
%%
We realized this fact a long time ago. We did not recognized, however, the
significance of the derived center of mass structure.  Only recently, when
the true soft mode of PMN was observed directly by neutron
scattering at high temperature,\cite{gehring_2001} did we recognize the
identity of the diffuse structure as a condensed soft mode.  Then the idea of
the phase shift followed naturally.
%%
%%}
%%
%% Rev.1 (end)
%%
But one important question still remains: {\em what is the microscopic origin
of this phase shift?} It should be emphasized that the direction of the phase
shift is not arbitrary as is clear from Fig.~\ref{FIG:Lattice}. Our model
requires that the phase shift be parallel, not antiparallel, to the
polarization resulting from the atomic displacement.  Under zero external
electric field, as is the case here, the polarization direction of a single PNR
can be parallel to any symmetric direction of $\langle 111 \rangle$,
i.e., $[111]$, $[\bar{1}11]$,$\cdots$,
$[\bar{1}\bar{1}\bar{1}]$.\cite{note1} Therefore, the phase-shifted
directions of the PNR are microscopically aligned to their
polarizations, although they appear macroscopically random.  Then,
{\em why is the phase shift of the PNR parallel to their polarization?}

We speculate that microscopic inhomogeneities in the site-occupancy of
the Mg$^{2+}$ and Nb$^{5+}$ cations produce a local electric field
gradient that determines the direction of polarization of the PNR.  In
other words, a nanoregion exposed to a local field gradient becomes a
{\em polar} nanoregion with a polarization parallel to the gradient.
Once a PNR is formed, its polarization starts to interact with the
local field gradient, and may result in shifting the PNR in order to
compensate for the Coulomb interaction with the lattice distortion due
to the phase shift.  The phase shift in this case is parallel to the
local field gradient, and thus to the polarization of the PNR.  Since
the microscopic inhomogeneities associated with the spatial
distribution of the Mg$^{2+}$ and Nb$^{5+}$ cations causes random
charged domains, the local field gradient is random.  Therefore, the
polarizations of the PNR should also be random.  However, as explained
above, the phase shift of a particular PNR is always parallel to its
polarization.  These descriptions are still qualitative, and need to
be tested experimentally.  It is particularly important to study the
microscopic properties of the PNR, such as their correlation lengths
and textures as a function of temperature as well as electric field.
We also need to understand the chemical inhomogeneity of Mg$^{2+}$ and
Nb$^{5+}$ microscopically.

Finally, our model should help to resolve some of the conflicting
x-ray diffuse scattering interpretations.
\cite{vakhrushev_1996,you_1997,takesue_preprint} The diffuse intensity
around (200) is only accidentally zero when measured with neutrons,
but it should be non-zero when measured with x-rays because of the
differences between the neutron nuclear scattering lengths and the
x-ray atomic scattering factors.

In summary, we have studied the diffuse scattering of the relaxor
Pb(Mg$_{1/3}$Nb$_{2/3}$)O$_{3}$ over a wide temperature range with
neutron scattering techniques.  We have confirmed that the relative
intensities are consistent with previous reports and that the diffuse
intensity starts increasing below the Burns temperature $T_{d}$.  We
have revisited the soft transverse-optic mode, the existence of which
has been recently confirmed, and examined the inconsistency between
the soft phonon and the diffuse intensities.  We have proposed a new
concept, the ``phase-shifted condensed soft mode,'' which naturally
explains the inconsistency as well as the microscopic origin of
the formation of the polar nanoregions.

\begin{acknowledgments}
  We would like to thank W.~Chen for his valuable help in the growth
  of the PMN crystal, Y.~Fujii, D.~Neumann, B.~Noheda, K.~Ohwada, H.~You, and
  S.~Vakhrushev for stimulating discussions, and especially N.~Takesue
  for sharing with us his recent x-ray scattering results on PMN.
  This work was supported mainly by the U.\ S.\ -- Japan Cooperative
  Research Program on Neutron Scattering between the U.\ S.\ 
  Department of Energy and the Japanese MONBU-KAGAKUSHO, and in part
  by a Grant-In-Aid for Scientific Research from the MONBU-KAGAKUSHO.
  We also acknowledge financial support from the U.\ S.\ DOE under
  contract No.\ DE-AC02-98CH10886, and the Office of Naval Research
  under Grant No.\ N00014-99-1-0738.  Work at the University of
  Toronto is part of the Canadian Institute for Advanced Research and
  is supported by the Natural Science and Engineering Research Council
  of Canada.  We acknowledge the support of the NIST Center for
  Neutron Research, the U.\ S.\ Department of Commerce, for providing
  the neutron facilities used in the present work.
\end{acknowledgments}

%
% now the references. delete or change fake bibitem. delete next three
%   lines and directly read in your .bbl file if you use bibtex.


\begin{references}
\bibitem{note1} Note that the polarization along $\langle 111 \rangle$
  is a superposition of three $\langle 100 \rangle$ components.  We
  have chosen only one of them, and have described our model along
  that direction.  However the more general case of $\langle 111
  \rangle$ is easily derived from this.
\end{references}
\end{document}